\documentclass[final,5p,times,twocolumn]{revtex4-1}

\usepackage{graphicx}
\usepackage{amssymb}
\usepackage{amsmath}
\usepackage{color}
\usepackage{lineno}
\usepackage{bigstrut}
\usepackage{multirow}
\usepackage{setspace}
\usepackage[exponent-product=\cdot,scientific-notation=false,separate-uncertainty=false,per-mode=symbol]{siunitx}

\usepackage{lineno}
\usepackage{booktabs}

\begin{document}


\title{Shape- and element-sensitive reconstruction of periodic nanostructures with grazing incidence X-ray fluorescence analysis and machine learning}





\author{A. Andrle}
\email{anna.andrle@ptb.de}
\affiliation{Physikalisch-Technische Bundesanstalt (PTB), 
Abbestr. 2-12, 10587 Berlin, Germany}

\author{P. H\"{o}nicke}
\affiliation{Physikalisch-Technische Bundesanstalt (PTB), 
Abbestr. 2-12, 10587 Berlin, Germany}

\author{G. Gwalt}
\affiliation{Helmholtz Zentrum Berlin für Materialien und Energie (HZB), Department Optics and Beamlines, Albert-Einstein-Str. 15, 12489 Berlin, Germany}

\author{P.-I. Schneider}
\affiliation{JCMwave GmbH, Bolivarallee 22, 14050 Berlin, Germany}
\affiliation{Zuse Institute Berlin, Takustrasse 7, 14195 Berlin, Germany}

\author{Y. Kayser}
\affiliation{Physikalisch-Technische Bundesanstalt (PTB), 
Abbestr. 2-12, 10587 Berlin, Germany}

\author{F. Siewert}
\affiliation{Helmholtz Zentrum Berlin für Materialien und Energie (HZB), Department Optics and Beamlines, Albert-Einstein-Str. 15, 12489 Berlin, Germany}


\author{V. Soltwisch}
\affiliation{Physikalisch-Technische Bundesanstalt (PTB), 
Abbestr. 2-12, 10587 Berlin, Germany}

\begin{abstract}
The characterization of nanostructured surfaces with sensitivity in the sub-nm range is of high importance for the development of current and next generation integrated electronic circuits. Modern transistor architectures for e.g. FinFETs are realized by lithographic fabrication of complex, well ordered nanostructures. Recently, a novel characterization technique based on X-ray fluorescence measurements in grazing incidence geometry has been proposed for such applications. This technique uses the X-ray standing wave field, arising from an interference between incident and the reflected radiation, as a nanoscale sensor for the dimensional and compositional parameters of the nanostructure. The element sensitivity of the X-ray fluorescence technique allows for a reconstruction of the spatial element distribution using a finite-element method. Due to a high computational time, intelligent optimization methods employing machine learning algorithms are essential for a timely provision of results. Here, a sampling of the probability distributions by Bayesian optimization is not only fast, it also provides an initial estimate of the parameter uncertainties and sensitivities. 
The high sensitivity of the method requires a precise knowledge of the material parameters in the modeling of the dimensional shape provided that some physical properties of the material are known or determined beforehand. The unknown optical constants were extracted from an unstructured but otherwise identical layer system by means of soft X-ray reflectometry. The spatial distribution profiles of the different elements contained in the grating structure were compared to scanning electron and atomic force microscopy and the influence of carbon surface contamination on the modeling results were discussed.
\end{abstract}

\maketitle

\section{Introduction}
Since nanotechnology and thus nanostructures of different kind are relevant in many areas of science and technology, metrology techniques that can support design, research and fabrication of such nanostructures are of high importance. Especially in the semiconductor industry, which is probably the most popular field of application for nanotechnology as well as a very strong driver for research in this field, complex 2D and 3D nanostructures with feature sizes in the single-digit nanometer regime~\cite{S.Natarajan2014,markov_limits_2014} are employed in order to keep Moore's law~\cite{more_1965} alive. The performance of these nanostructures crucially depends on a well-controlled fabrication, both in terms of targeted dimensional parameters and 3D element compositions (e.g. dopant distributions). Thus, there is a strong need for metrology techniques which allow to characterize these parameters with sufficient sensitivity~\cite{Orji_2018}.

Typical analytical or dimensional techniques used in this context are scanning and transmission electron microscopy (SEM, TEM)~\cite{raymond_sub-nanometer_2011,erni_atomic-resolution_2009}, atomic force microscopy (AFM)~\cite{ukraintsev_role_2005} and techniques that also address elemental distributions such as secondary ion mass spectroscopy (SIMS) ~\cite{Franquet_2016}, atom probe tomography (APT) ~\cite{vandervorst_dopantcarrier_2014} and energy-dispersive X-ray spectroscopy (EDX) combined with scanning electron microscopy (STEM)~\cite{mertens_vertically_2017}. All these techniques have different advantages and disadvantages regarding sample preparation and consumption, achievable spatial resolution, required duration and other experimental parameters (e.g.~tip sizes). Optical metrology based on light-matter interaction has a significant advantage in terms of measurement speed (with respect to slow techniques, e.g.~APT and STEM) and the ability to statistically measure large areas in contrast to scanning techniques. Optical reflectometry also known as optical critical dimension (OCD)~\cite{silver_fundamental_2007} metrology is still used and continuously improved despite the resolution limits that have been reached. In order to keep pace with shrinking structures, intensive research is being carried out to reduce the wavelength of the employed radiation and to increase the sensitivity via the dispersion of the periodic nanostructured surface. This is called deep ultraviolet (DUV) or extreme ultraviolet (EUV) scatterometry and can be extended into the X-ray spectral range. In X-ray scattering techniques, a distinction is made between measurements in transmission, also known as critical dimension small angle X-ray scattering (CDSAXS)~\cite{Sunday_2015}, and reflection mode known as grazing incidence small angle X-ray scattering (GISAXS)~\cite{levine_grazing-incidence_1989, renaud_probing_2009}. Both techniques have already shown that they allow for the dimensional reconstruction of nanostructures with an uncertainty in the sub-nm range~\cite{Soltwisch2017,FernandezHerrero2020}, but both require a special sample thinning or sample design, which limits the application possibilities~\cite{pfluger_extracting_2020}. Usually, these techniques also employ rather high photon energy X-rays, which limits the optical contrast between different materials within the investigated nanostructures.

In this paper, we employ the grazing incidence X-ray fluorescence analysis (GIXRF)~\cite{D.K.G.DeBoer1995} method and soft X-rays ($<1$keV) instead to analyze the nanostructured surface. This non-destructive technique allows reconstructing the composition of the sample in terms of both their dimensional properties as well as the distribution of different elements, since the emitted X-ray fluorescence (XRF) is characteristic for each element. GIXRF uses the X-ray standing wave (XSW) field ~\cite{Bedzyk_1989,Golovchenko1982}, which results from the interference between the incident and the reflected X-ray beam as a nanoscale sensor. The intensity modulation inside the XSW field significantly influences the X-ray fluorescence intensity of an element depending on its spatial position within the electromagnetic field distribution. Recent studies have shown the potential of the GIXRF technique for the dimensional and compositional nanometrology of periodic 2D~\cite{V.Soltwisch2018} and 3D~\cite{Nikulaev2019,honicke_grazing_2020} nanostructures. 

In this work, we further develop this approach towards being a reliable metrology technique by employing a combined element sensitive reconstruction for the fluorescence signals of oxygen, nitrogen and in parts also carbon from within a silicon nitride grating structure. The integration of machine learning (ML) techniques, such as the Bayesian optimization~\cite{shahriari2016taking} algorithm based on Gaussian processes in combination with a finite-element Maxwell solver~\cite{schneider2019benchmarking} allows to control the computational modeling effort and to derive first estimates of the uncertainties of the parameterized nanostructure. For an even increased reliability, we are experimentally determining the optical constants of the employed materials instead of using tabulated data as these are often not realistic for nanolayers and nanostructures~\cite{zhang_thickness-dependence_2012}. Finally, we compare and validate the results against AFM and SEM cross sections. 

\section{Experimental}
\begin{figure*}[ht]
\centering
\includegraphics[width=0.85\textwidth]{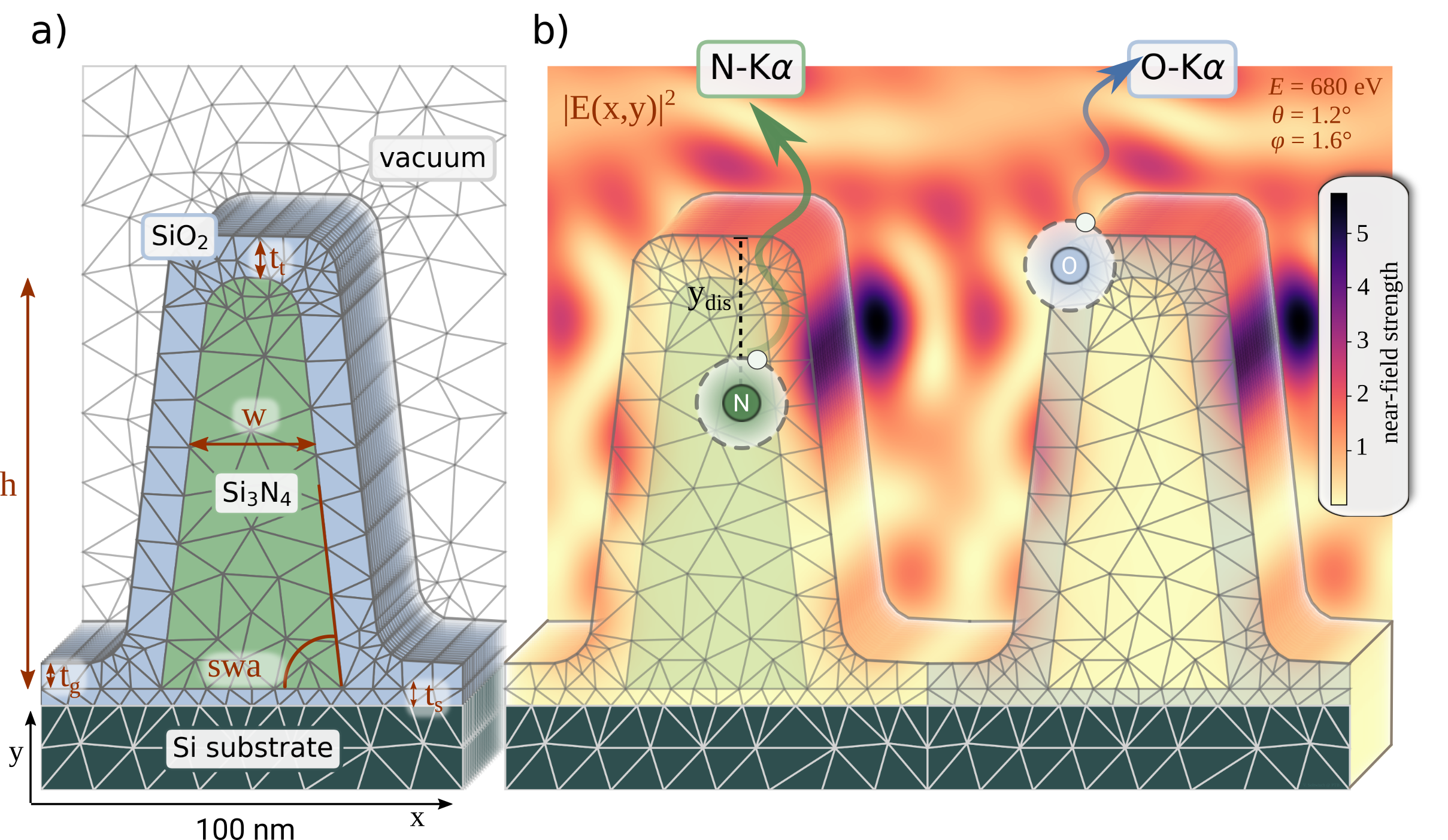}
	\caption{a) Cross-section with the finite-element mesh-grid showing the layout used for the simulation. The height $h$, the width $w$, the sidewall angle $swa$, the oxide layer thicknesses in the groove $t_g$ and the on the grating line $t_t$ were optimized as independent parameters. The oxide layer on the substrate $t_s$ was kept constant during the optimization.  b) The calculated electric field strength inside and outside the structure is shown for $\theta = 1.2^{\circ}$ and $\varphi = 1.6^{\circ}$. For the nitrogen and oxygen fluorescence the electric field strength is integrated and the reabsorption is calculated with the distance y$_{dis}$ the photon has to travel to leave the structure.}
\label{fig:fem_mesh}
\end{figure*}

As an example of two-dimensional nanostructures, a lithographically patterned silicon nitride grating on a silicon substrate was investigated. It was manufactured by means of electron beam lithography (EBL) at the Helmholtz-Zentrum Berlin. The nominal pitch of the grating is $p = 100$ $\si{\nm}$, the nominal height is $h = 90$ $\si{\nm}$ and the nominal line width of the sample is $w = 50$ $\si{\nm}$. 
For the manufacturing of the gratings, a silicon substrate with a 90 nm-thick Si$_3$N$_4$ layer was used. ZEP520A, a positive resist (organic
polymer), was spin coated on the substrate and developed with a Vistec EBPG5000+ e-beam writer, operated with an electron acceleration voltage of 100 kV. The grating was etched via reactive ion etching using CHF$_3$ and to remove the remaining resist an oxygen plasma treatment was applied.
A sketch of the cross-section is shown in Fig.~\ref{fig:fem_mesh} a). The total structured area of the grating was 1 mm x 15 mm and the sample area outside the patterned region consists of the originally deposited Si$_3$N$_4$ layer. Directly after fabrication, cross-section SEM images have been recorded on a sister sample.

We performed GIXRF and X-ray reflectometry (XRR) measurements in PTB's laboratory~\cite{B.Beckhoff2009c} at the BESSY II electron storage ring using the plane-grating monochromator (PGM) beamline~\cite{F.Senf1998} for undulator radiation. The sample was mounted in an ultrahigh-vacuum (UHV) measurement chamber~\cite{J.Lubeck2013}, where a 9-axis manipulator allows for an accurate sample alignment with respect to the direction of incident X-ray beam. The incidence angle $\theta$ is defined as the angle between the X-ray beam and the sample surface. The azimuthal angle $\varphi$ is defined as the angle between the incident beam and a plane, which is normal to the sample surface and parallel to the direction of the grating lines such that $\varphi = 0^{\circ}$ is defined as the orientation parallel to the plane of incidence. Both sample rotation axes can be aligned with an uncertainty below 0.01$^{\circ}$.

As we have employed radiometrically calibrated X-ray fluorescence (XRF) instrumentation, we can perform reference-free GIXRF~\cite{Hoenicke2019} and gain a quantitative access to the elemental mass depositions present on the sample~\cite{Beckhoff2008}. At each angular position for $\theta$ and $\varphi$, a fluorescence spectrum is recorded with a calibrated silicon drift detector (SDD)~\cite{F.Scholze2009} and the incident photon flux is monitored by means of a calibrated photodiode. The GIXRF-measurements were performed at an incident photon energy of $E_i = 680$ $\si{\eV}$, allowing for the excitation of N-K${\alpha}$ as well O-K${\alpha}$ fluorescence radiation, which mainly originates from the surface oxide layer on the grating structure. In Fig.~\ref{fig:GIXRF} the obtained and normalized N-K${\alpha}$ a) and O-K${\alpha}$ b) fluorescence intensities for different angles $\theta$ and $\varphi$ are shown.

In addition, we have performed XRR experiments on the non-structured Si$_3$N$_4$ layer next to the grating at the same photon energy ($E_i = 680$  $\si{\eV}$). From this we can determine the optical constants of the SiO$_2$ layers and the Si$_3$N$_4$ layer, which are expected to be more reliable in the soft X-ray spectral range than using only tabulated data as in~\cite{andrle_anisotropy_2020}.

For an independent validation of the dimensional GIXRF reconstruction results, additional AFM measurements were performed with an Nanosurf Nanite 25x25. The sample was measured under tapping mode condition and a standard pyramidal shaped silicon probe with a tip radius $<10$ $\si{\nm}$ was used, as it is commonly applied for the inspection of diffraction gratings~\cite{siewert_gratings_2018}. The AFM-probe is characterized by a resonance frequency of $190$ $\si{\Hz}$ and force constant of $48$~$\si{\N}/\si{\m}$. The inspected sample area was $500$x$500$~$\si{nm}^2$ in size covering about 5 grating lines. The height profile, obtained by averaging two in juxtaposition located AFM line profiles, is shown in Fig.~\ref{fig:AFM}.

\section{Simulation and optimization of fluorescence intensities}
The GIXRF signals are directly related to the XSW field intensity distribution which, besides the incident photon energy and the incidence and azimuthal angles, depends on the shape and material composition of the illuminated nanostructured surface. To reconstruct these sample features from the experimental data, we applied a finite-element method (FEM) based forward calculation of the XSW and optimized the structural parameters to reproduce the experimental data. 

 In Fig.~\ref{fig:fem_mesh}, the basic principle of the FEM procedure is shown. The mesh grid and the model parameters are displayed in Fig.~\ref{fig:fem_mesh} a). The FEM solver calculates the electric near-field for a given structure (Fig.~\ref{fig:fem_mesh} b)). The amount of fluorescence photons generated at a given coordinate depends on the local electric near-field intensity ($E(x,y)$) and the compositional and fundamental parameters of the respective material (mass fraction of the fluorescent element in the material $W_k$, photo-ionization cross section for the incident photon energy $\tau(E_i)$ and the fluorescence yield $\omega_k$, taken from databases~\cite{T.Schoonjans2011} or dedicated experiments~\cite{P.Hoenicke2016a}). These photons can be reabsorbed on the path through the sample towards the detector (Fig.~\ref{fig:fem_mesh} b)) with a probability depending on the materials mass attenuation coefficient $\mu(E_f)$ for the fluorescence photon energy $E_f$, the density $\rho$ of the material and the distance to the surface of the nanostructure $y_{dis}(x,y)$. Eventually, the fluorescence photon will be detected with a given detection efficiency $\epsilon(E_f)$ if it is arriving within the effective solid angle of detection $\Omega$. The overall emitted fluorescence intensity of course also depends on the incident photon flux $N_0$. 
  
Thus, the measured emitted fluorescence intensity $\Phi(\theta,\varphi,E_i)$ (derived from the detected count rate $F(\theta,\varphi,E_i)$) can be modeled using the integration over the full area of the fluorescent material of the nanostructure as described by the modified Sherman equation~\cite{V.Soltwisch2018, honicke_grazing_2020}:

\begin{widetext}
\begin{equation}
\Phi(\theta,\varphi,E_i) = \underbrace{ \frac{4\pi\sin\theta}{\Omega}\frac{F(\theta,\varphi,E_i)}{N_0\epsilon(E_f)} }_{\substack{I_{exp}}}= \underbrace{\frac{W_k \rho \tau(E_i) \omega_k }{\sum dx} \cdot \sum_{x}\sum_{y}  |E(x,y)|^2 \cdot \exp\left[-\rho\mu(E_f)y_{dis}(x,y)\right] dx dy}_{\substack{I_{model}}}\textrm{.} 
\label{eq:2Dsherman}
\end{equation}
\end{widetext}

The equation~\ref{eq:2Dsherman} is applied for nitrogen and oxygen fluorescence and has been implemented directly in the Maxwell solver to eliminate errors due to conversion to a regular cartesian grid. The calculated emitted fluorescence intensities $I_{model}$ are then compared against the experimental value $I_{exp}$. By using a global optimization algorithm such as Bayesian optimization (BO)~\cite{shahriari2016taking,schneider2019benchmarking}, an optimal set of the model parameters can then be determined. BO uses a stochastic model, a Gaussian process, of an unknown objective function to be minimized in order to determine promising parameter values~\cite{shahriari2016taking,Garcia-Santiago2018}. In a previous work~\cite{schneider2019benchmarking, Andrle_2019}, it was shown that the BO performs much better than other meta-heuristic optimization approaches with respect to the computing time needed to find the global minimum. Since BO considers all previous function evaluations, it can be more efficient than other meta-heuristic global optimization strategies~\cite{schneider2019benchmarking} and local optimization strategies~\cite{schneider2019using}. This is a crucial benefit here, as one model calculation takes several minutes on a standard desktop computer. Here, we use an implementation of BO which is part of the JCMsuite software package~\cite{jcm_web:accuracy_2020}.

The error function $\chi^2$ 
\begin{equation}
\chi^2(\vec{gp}) = \sum_{\theta,\varphi}{\frac{(I_{exp}(\theta,\varphi) - I_{model}(\vec{gp},\theta,\varphi))^2 }{\sigma_N^2(\theta,\varphi)}}
\label{eq:chi}
\end{equation}
was minimized with respect to the different model parameters of $\vec{gp}$ described earlier (see Fig.~\ref{fig:fem_mesh}~a)) and model errors for the nitrogen $\epsilon_N$ and the oxygen fluorescence signals $\epsilon_O$). These model error parameters are introduced to consider potential errors e.g.~the uncertainty of the employed atomic fundamental parameters or the influence of a thin surface contamination layer into account. Even though material dependent parameters like optical constants or material densities deviate most likely from the tabulated bulk data for the grating materials (SiO$_2$ and Si$_3$N$_4$), we do not include these as free parameters in the model. Indeed, as this would drastically increase the number of model parameters and thus prolong the necessary calculation times, we rather determine these parameters separately by XRR (described in the next section). 
$\sigma_N$ is the calculated experimental error consisting of an error estimation for the effective solid angle of detection $\sigma_{\Omega}(\theta)$, the error contributions originating from counting statistics $\frac{\sqrt{F}}{F}$ for the respective fluorescence line as well as for the spectra deconvolution $\sigma_{num}$.
\begin{equation}
\sigma_N(\theta)^2 = \left( \frac{\sqrt{F(\theta)}}{F(\theta)} \right)^2 + \sigma_{\Omega}(\theta)^2  + \sigma_{num}(\theta)^2 
\end{equation}

By inverting the Hessian matrix ${\bf H}_{k j}$~\cite{henn_maximum_2012} of the error function,
\begin{equation}
{\bf H}_{k j} = \frac{\partial^2 \chi^2(\vec{gp})}{\partial gp_j \partial gp_k} 
\end{equation}
at the minimum of $\chi^2$ it is possible to determine the confidence intervals $\left(\bf H\right)^{-1}$ of the model parameters, if they are gaussian distributed as assumed.

The standard deviation or noise level at the global minimum parameter set is defined as ${\rm STD} = \sqrt{\frac{\chi^2}{\rm DOF}}$, where ${\rm DOF} = N-M$ is the difference between the number of measurement points $N$ and the number of free model parameters $M$ and thus the degrees of freedom. 

The model parameter confidence interval can then be calculated by
\begin{equation}
 \sigma_{\vec{gp}} = {\rm STD}\sqrt{{\rm diag}(\bf \left(\bf H\right)^{-1})}
\label{eq:STD}
\end{equation}

Based on the Gaussian process model of $\chi^2(\vec{gp})$ we determine an estimate of the Hessian matrix by computing all second derivatives of the Gaussian process model of $\chi^2$ at the minimum and can then estimate the error covariance matrix $\left(\bf H\right)^{-1}$ and the parameter confidence interval. 

\section{Results and Discussion}
\subsection{Validation of the optical material parameters}
\begin{figure}[ht]
\centering
\includegraphics[width=0.45\textwidth]{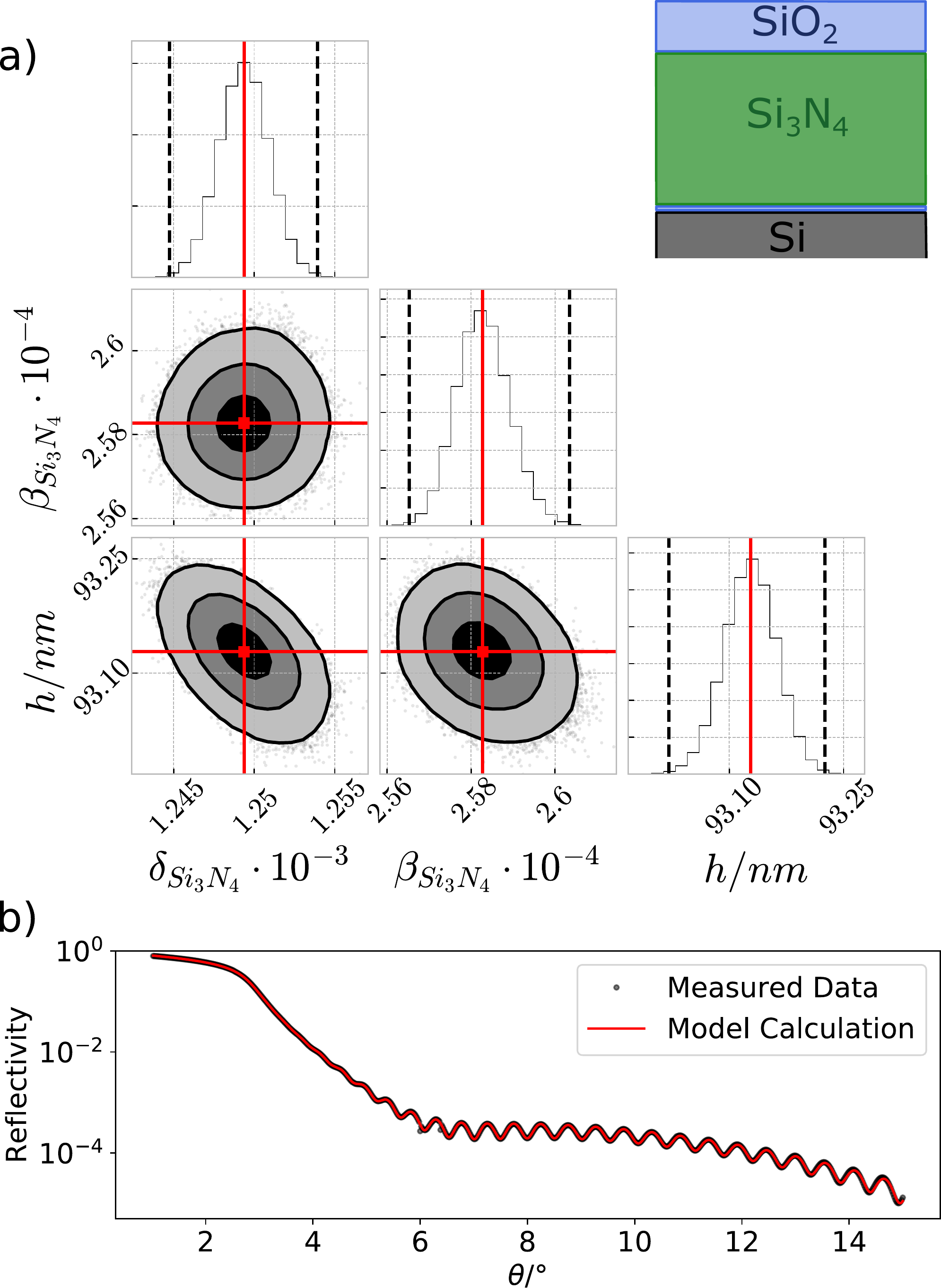}
	\caption{a) The posterior distributions for relevant parameters $\delta_{Si_3N_4}$, $\beta_{Si_3N_4}$ and the $Si_3N_4$ thickness $h$ obtained from the MCMC sampling. The red line marks the mean of the distribution and the dotted black lines in the histogram indicates a 3$\sigma$ interval. In the top right corner, a sketch of the used layer stack is shown.\\
	b) Comparison of the experimental data (black stars) and the model calculation (red line) as obtained by the MCMC.}
\label{fig:corner_XRR}
\end{figure}

From the angle-dependent measurement of reflection intensities on a layered system, information about layer thicknesses, densities or even their optical constants can be obtained~\cite{chason_thin_1997}. 
Fig.~\ref{fig:corner_XRR} b) shows the experimental data in comparison with the best simulation. The high frequency oscillation visible in the XRR curve, is a clear indication for a multi-layer system. Due to the native oxide layer of the Si substrate and the removal of the photo resist, resulting in an oxidized surface on the Si$_3$N$_4$ layer~\cite{kennedy_oxidation_1999}, we apply a three layer model for the XRR simulation (as shown in the inset of Fig.~\ref{fig:corner_XRR} a)). Since the calculation of the reflectivity for a 1D layer system is several orders of magnitude faster than the FEM based 2D GIXRF modeling, statistical analysis methods of the posterior distributions can be used for a large number of parameters such as layer thickness, roughness and optical constants. We applied the Markow Chain Monte Carlo method (MCMC)~\cite{emcee} to determine the individual parameter uncertainties and to resolve possible inter parameter correlation effects~\cite{Haase2016}. In Fig.~\ref{fig:corner_XRR} a) the posterior distribution determined in this procedure is shown as projections of the refractive index $n(Si_{3}N_{4}) = 1-\delta+i\beta$ and thickness $h$ of the $Si_{3}N_{4}$ layer. The almost perfect gaussian-like shape of the distributions, which is also present in all other parameters, allows the determination of uncertainties directly from the measurements. The relative uncertainties of the experimental data reconstructed with a linear error model ($ax+b$)~\cite{henn_maximum_2012} are with $(0.6\pm 0.2)\%$ exactly in the expected range.

From this modeling we derived optical constants for the top $SiO_2$ ($\delta$ = $(8.73 \pm 0.03) 10^{-4}$, $\beta$ = $(2.58 \pm 0.04) 10^{-4}$) and the $Si_3N_4$ ($\delta$ = $(12.59 \pm 0.05) 10^{-4}$, $\beta$ = $(2.58 \pm 0.02) 10^{-4}$) layers. By comparing the experimentally determined optical constants with tabulated Henke data~\cite{henke_x-ray_1993}, one can estimate the densities of the respective materials and their deviation from the respective bulk densities. For $Si_3N_4$, a relative density of $(0.89 \pm 0.01)$ and for $SiO_2$ a relative density of $(0.79 \pm 0.01)$ was found. This is in line with the already observed material density reduction discussed in Ref.~\cite{V.Soltwisch2018}. The reduced densities as well as the experimental optical constants are used for the GIXRF reconstruction.

\subsection{GIXRF reconstruction results}
\subsubsection{Virtual Experiment}
\begin{figure}[tbp]
\centering
\includegraphics[width=0.49\textwidth]{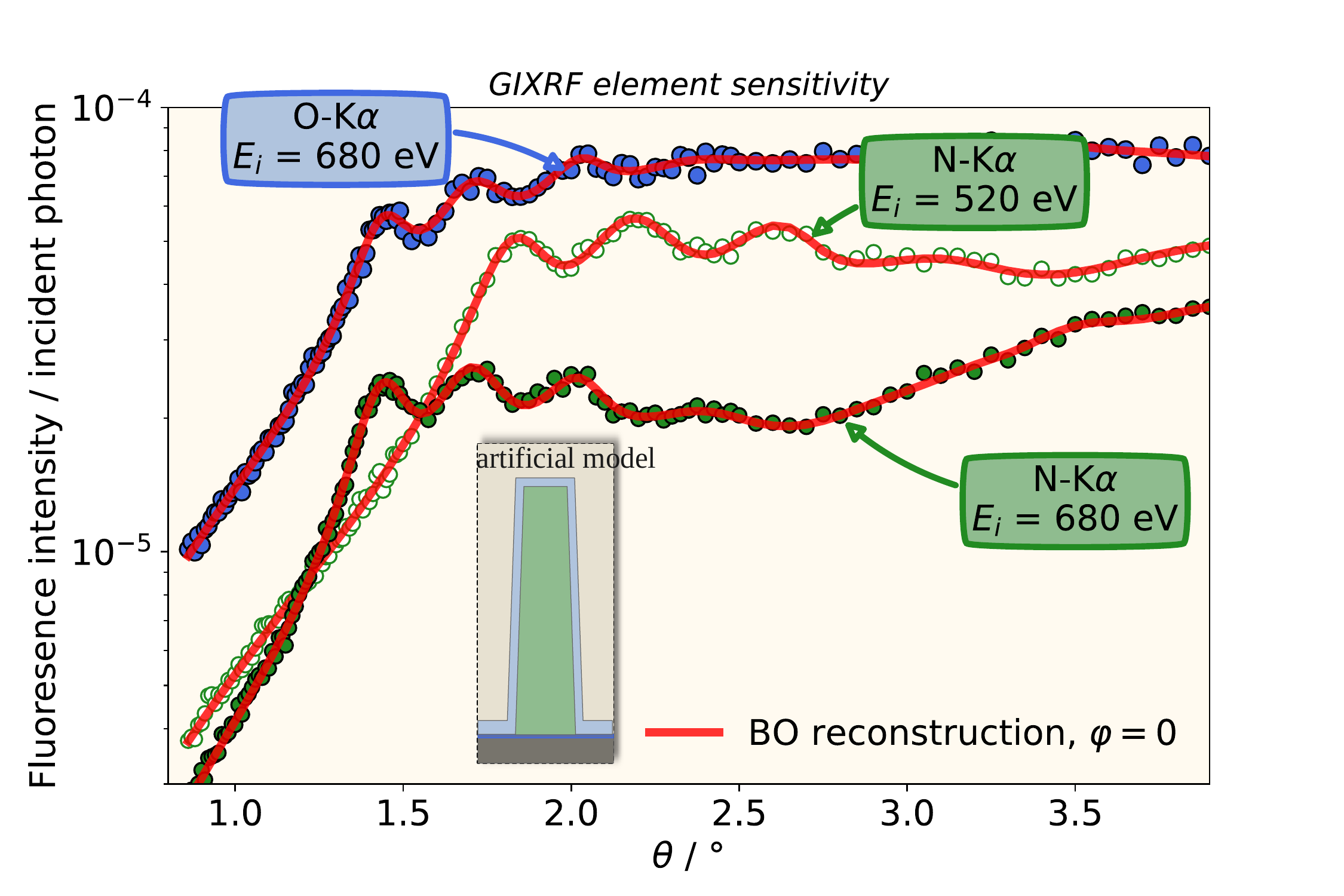}
	\caption{Comparison of the expected artificial disturbed simulated N-K$\alpha$ (green points) or O-K$\alpha$ (blue points) fluorescence intensities for the different excitation energies and the corresponding BO reconstruction. Clearly visible is how the first peak in the nitrogen signal shifts with increasing energy to smaller incident angles.}
\label{fig:GIXRF_simu}
\end{figure}

Before we apply the reconstruction model to real experimental data, we apply it to an artificial dataset in order to test the reconstruction method and to study, whether an increased incident photon energy, capable of exciting also oxygen fluorescence signal, is beneficial. In our previous study~\cite{V.Soltwisch2018}, measurement data at $E_i = 520$~$\si{\eV}$ was analyzed and an indirect sensitivity to the surface oxide layer was found even though no fluorescence signal originating from it was used for the reconstruction. The indirect sensitivity was merely due to the attenuation behavior of the oxide layer and thus, we expect an increased sensitivity if also a direct signal originating from it is used. For this purpose, 
we generate artificial experimental data by calculating model curves using the reconstruction model for a given set of parameters  and the two incident photon energies of $E_i = 680$~$\si{\eV}$ and $E_i = 520$~$\si{\eV}$. To mimic experimental noise, we apply a Gaussian disturbance with a width of 3\% (see  Tab.~\ref{tab:results_GIXRF_sim}). 
By reconstructing the artificial datasets, we can now analyze the influence of the increased incident photon energy on the reconstruction results and their confidence intervals without the influence of any experimental error contributions.
In Fig.~\ref{fig:GIXRF_simu} the corresponding artificial GIXRF curves and reconstruction results are displayed. They match the artificial data well for both photon energies and characteristic features, e.g. the first local maxima are very well retrieved. From the BO reconstruction, we are able to calculate the confidence intervals of the reconstructed parameters and, as shown in Tab.~\ref{tab:results_GIXRF_sim}, they agree well with the initial parameters within the derived confidence interval.

From a comparison of the determined confidence intervals for the two different configurations (Configuration A, $E_i = 520$~$\si{\eV}$, only the nitrogen signal is modeled and Configuration B, $E_i = 680$~$\si{\eV}$, both nitrogen and oxygen signals are modeled, see Tab.~\ref{tab:results_GIXRF_sim}), the positive influence of also considering the oxygen fluorescence signal is obvious. Especially for the height and the groove oxide thickness, the achieved confidence intervals are significantly smaller for configuration B. Nevertheless, also for configuration A, the angle-dependent nitrogen fluorescence contains all relevant information about the dimensional properties of the nanostructure and even the surface oxide layer, as already pointed out in our earlier work~\cite{V.Soltwisch2018}.  

\begin{table}
\caption{The values of the geometrical parameters of the synthetic data from the GIXRF BO reconstruction with the model parameter confidence intervals for $E_i = 520$~$\si{\eV}$ (Configuration A, only the nitrogen signal is modeled) and $E_i = 680$~$\si{\eV}$ (Configuration B, both nitrogen and oxygen signals are modeled). The pitch was set to $p = 50$~$\si{\nm}$. (height $h$, width $w$, sidewall angle $swa$, oxide layer in the groove $t_g$ and oxide layer on the grating line $t_t$)}
\begin{tabular}{c|c|c|c|c}
Parameter& Intial & \multicolumn{2}{c}{Config.} & Conf. Interval  \\ 
Name & Value& A & B & Ratio $\frac{B}{A}$\\
\hline
$h/\si{\nm}$  & $90$ &   $89.4(6)$   & $90.3(4)$  & 0.67\\
$w/\si{\nm}$   &   $25$ &   $24.87(7)$    & $25.08(6)$ & 0.85 \\
$swa/^{\circ}$  &  $88$ &    $88.1(1)$      & $87.9(1)$ & 1.0\\
$t_t/\si{\nm}$ &    $3$&    $3.13(6)$     & $3.02(5)$  & 0.83 \\
$t_g/\si{\nm}$ &$5$&      $5.3(6)$       & $5.0(1)$  & 0.17 \\
$\epsilon_N$      &$1$&     $1.02(1)$      & $1.00(1)$ & 1\\
$\epsilon_O$      &$1$&     -             & $0.99(1)$&  
    
\label{tab:results_GIXRF_sim}
\end{tabular}
\end{table}

\subsubsection{Real experimental data}

\begin{figure}[tbp]
\centering
\includegraphics[width=0.45\textwidth]{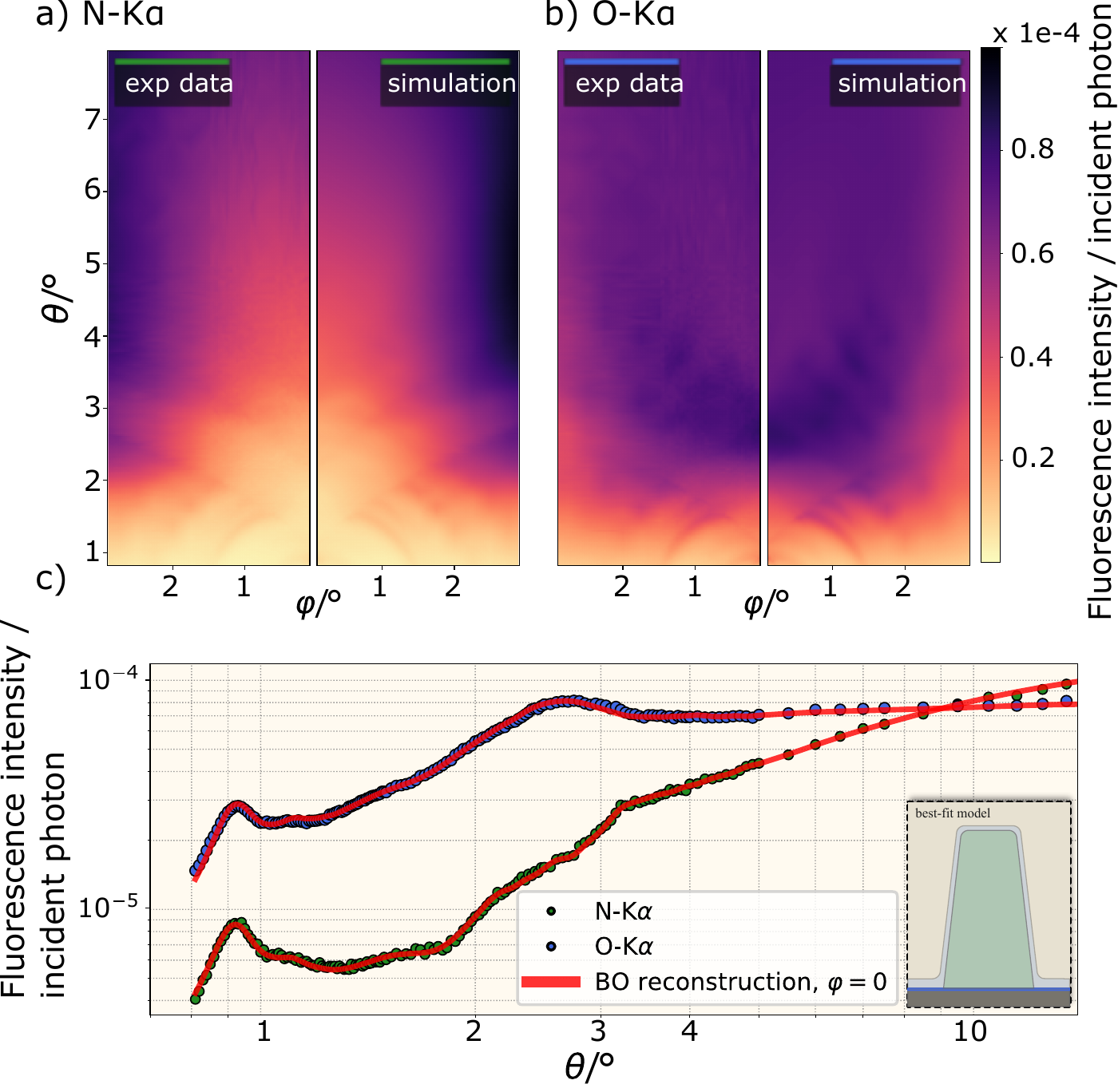}
	\caption{Comparison of the measured and simulated fluorescence maps for N-K a) and O-K b) based on the reconstructed parameter set.  
	c) Comparison of the experimental N-K$\alpha$ (green) or O-K$\alpha$ (blue points) fluorescence intensities for $\varphi = 0^\circ$ to the reconstructions from the Bayesian optimization (red lines).}
\label{fig:GIXRF}
\end{figure}

Using the same model and methodologies, also the real experimental data, measured according to configuration B, was modeled. A  comparison of the two experimental datasets and the resulting modeled data is shown in Fig.~\ref{fig:GIXRF}. The corresponding optimal parameters and confidence intervals are summarized in Tab.~\ref{tab:results_GIXRF}. Again, the confidence intervals of the structure parameters derived from the BO posterior distribution are very similar to those determined in the virtual experiment.

Similarly as already for the virtual experiment, a higher sensitivity for the line width is observed as compared to the height. This is expected to be a result of the relatively large line height with respect to the achieveable information depth in the soft X-ray regime. Thus, the nitrogen fluorescence radiation from the very bottom of the grating line does not contribute significantly to the overall observed signal. Nevertheless all calculated confidence intervals are in the sub-nm regime.

It should also be noted that the two modeling error parameters $\epsilon_N$ and $\epsilon_O$ deviate from unity within a range of 10~\%. This is the same magnitude one would expect the relative uncertainty of the employed fluorescence production cross sections (product of fluorescence yield and photo ionization cross section) to be in.

Systematic errors not taken into account may increase the final uncertainties. This problem is often called model error and refers to the whole physical model or virtual experiment that is applied and is not limited to the finite element model. The next section shows that these modeling uncertainties can have a significant impact on the reconstruction parameters. 
\begin{table}
\caption{The values of the geometrical parameters of the nanostructures from the GIXRF BO reconstruction with the model parameter confidence intervals (one sigma). The parameters are height $h$, width $w$, sidewall angle $swa$, oxide layer in the groove $t_g$ and oxide layer on the grating line $t_t$.
}

\begin{tabular}{c|c}
Parameter& Reconstructed   \\ 
Name & Value\\
\hline
$h/\si{\nm}$    & $97.5(5)$     \\
$w/\si{\nm}$    & $49.77(7)$     \\
$swa/^{\circ}$  & $83.54(9)$    \\
$t_t/\si{\nm}$  & $2.84(3)$      \\
$t_g/\si{\nm}$  & $5.82(9)$       \\
$\epsilon_N$       & $0.918(6)$     \\
$\epsilon_O$       & $1.059(7)$

\label{tab:results_GIXRF}
\end{tabular}
\end{table}

\subsubsection{Contamination}
\begin{figure}[tbp]
\centering
\includegraphics[width=0.45\textwidth]{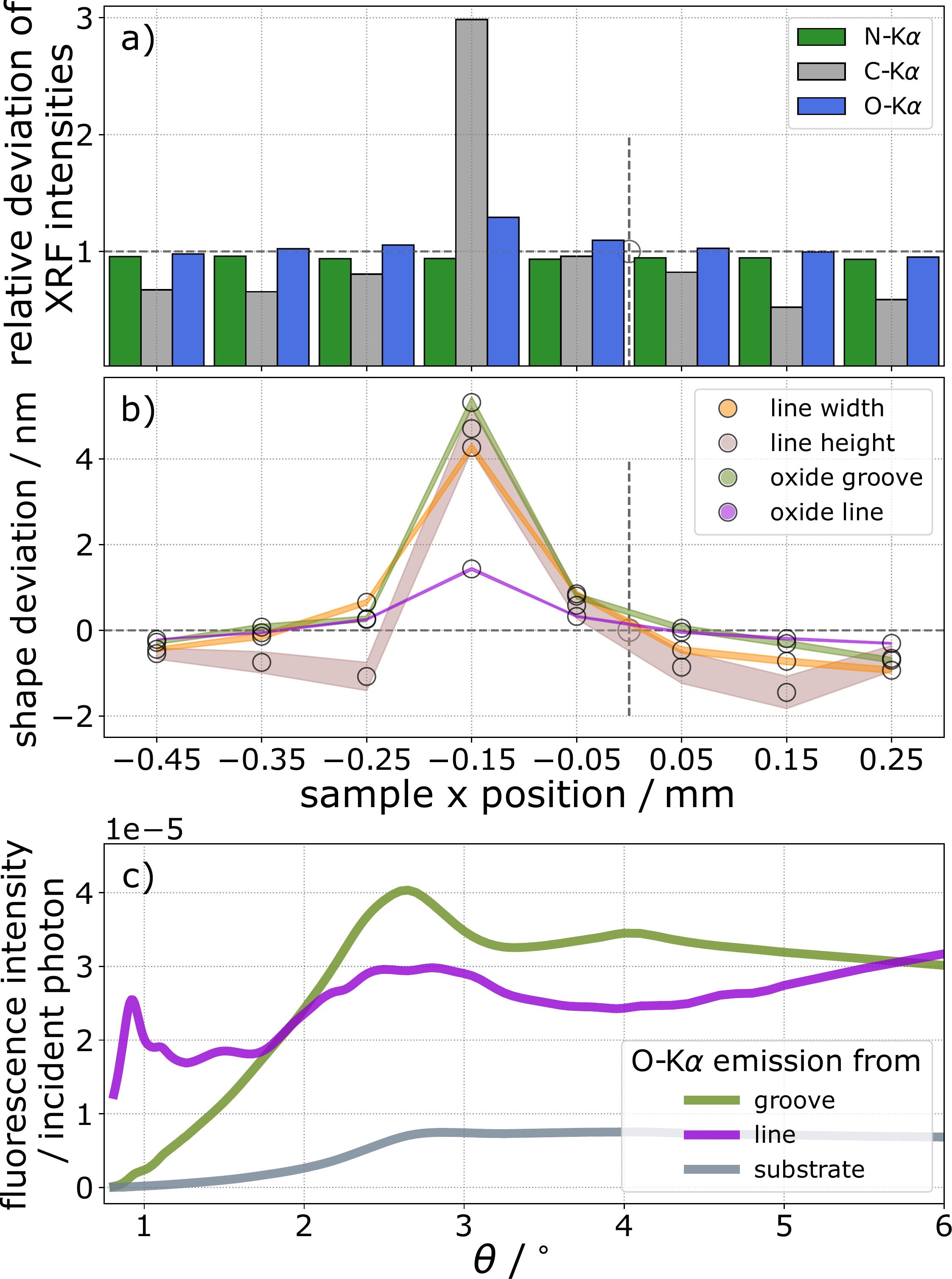}
	\caption{a) Here the C-K$\alpha$, N-K$\alpha$ and O-K$\alpha$ fluorescence intensity at $\theta=15^\circ$ where no XSW needs to be considered from different positions on the sample are compared. \\
	b) The reconstruction results for the width, the height, the thickness of the oxide in the groove and the thickness of oxide around the line  for the different position are plotted. \\
	c) The various contributions from the oxide from the line, the groove and the oxide layer of the silicon wafer to the total oxygen fluorescence are compared.
	}
\label{fig:GIXRF_PrX}
\end{figure}

In addition to oxygen and nitrogen signals in the fluorescence spectra, also carbon fluorescence was observed. A presence of carbon on the sample surface is likely as the sample is stored under normal ambient conditions. A lateral scan of the sample (Fig.~\ref{fig:GIXRF_PrX} a)) at a fixed incident angle $\theta = 15^{\circ}$ reveals that this carbonaceous contamination is not homogeneously distributed over the patterned area. In the center of the grating area ($x=-0.15$~$\si{\mm}$) a strong increase of the carbon signal and a slight increase of the oxygen signal  can be observed, whereas the nitrogen signal is practically constant.
In an earlier benchmark study, the sample under investigation was measured in various scatterometers and electron microscopes around the world. A punctual contamination of the grating surface by these techniques can therefore not be excluded. Here, we take advantage of this artificially created contamination layer to demonstrate the sensitivity and the influence of model errors.

At each lateral position shown in the figure, we performed a GIXRF angular scan (at $\varphi$ = 0$^\circ$) and performed the reconstruction including the confidence interval calculation without considering any contamination. 
Fig.~\ref{fig:GIXRF_PrX}~b) shows the obtained differences between the reconstructed model parameters at each position to the reference position at $x=0$  $\si{\mm}$ (position where the data shown in Fig.~\ref{fig:GIXRF} was taken). The model parameters at $-0.15$~$\si{\mm}$, which is the position with maximal carbon and oxygen contamination, differ clearly with respect to less contaminated areas. The difference is larger than the calculated confidence intervals from the reconstruction and they are also much larger than the expected lateral inhomogeneities of the sample. In addition, the nearly constant nitrogen fluorescence signal in X-direction clearly indicates a homogeneous overall amount of $Si_3N_4$, which is in contradiction to the larger cross-sectional area of the grating determined by the reconstruction from the GIXRF model. The reconstructed heights and widths increase by more than $4$~$\si{\nm}$. 

This behavior can be explained considering the XSW near field distribution inside the grooves: The slight increase of the oxygen due to the contamination signal can only be incorporated by increasing the oxide layer thicknesses. However, an increase of the oxide layer thickness weakens the penetrating field inside the $Si_3N_4$ and thus the emitted nitrogen fluorescence. To compensate for this, a larger grating cross section is reconstructed. One may think, that the reconstruction algorithm could circumvent this by simply increasing only the groove oxide layer thickness, which does not affect the nitrogen signal so much. But as shown in part c) of the figure, where the angular oxygen fluorescence signal contributions from the different parts in the structure model are shown, the groove oxide has a significantly different angular behavior as compared to the oxide on the grating line surface. Especially the features at about $1^\circ$ and $2.7^\circ$ are very distinct. For this reason, the reconstruction algorithm must increase both in order to account for the higher oxygen signal in the contaminated area throughout the angular range.

This shows how sensitive the GIXRF method is, but also how carefully the models have to be developed for a realistic uncertainty estimation. However, even if the model is not accurate enough, a spacial resolved reconstruction can reveal flaws of the model.


\subsection{Comparison with SEM and AFM}
For a validation of the dimensional and compositional parameters as derived from the GIXRF modeling without the thick carbon contamination, we have performed AFM measurements on the sample and use also SEM (on a witness sample). As AFM is an established method to map the surface topology of nanostructures with sensitivities down to the $\si{\nm}$ regime, we can use an AFM to verify the GIXRF reconstructed total line heights (difference between top surface and groove surface). From the AFM data, we determined a total line height of $98.4$~$\si{\nm}$. This is consistent with the GIXRF reconstructions within the uncertainties. Here, it should be noted that the AFM measurement is only representative for the small area of the grating where it was performed whereas the GIXRF result is averaged over a much larger area due to the elongated beam footprint. Due to the interplay of the tip shape with the nanostructure, other dimensional parameters as e.g.~line width or sidewall angle are not straightforward to be deduced~\cite{doi:10.1063/1.111578}.  

\begin{figure}[tbp]
\centering
\includegraphics[width=0.45\textwidth]{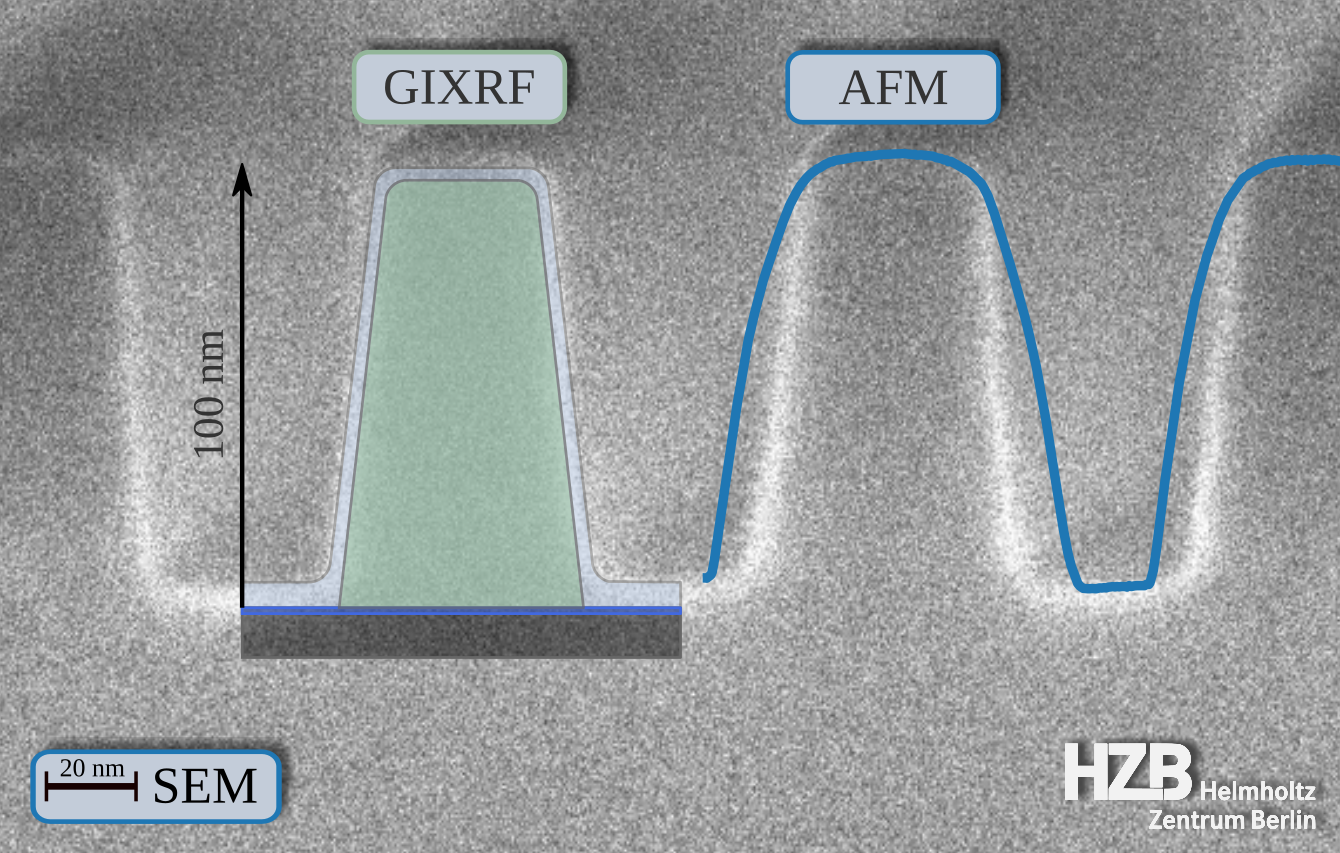}
	\caption{Comparison of the AFM data with the result of the GIXRF reconstruction and a SEM picture from a witness sample.}
\label{fig:AFM}
\end{figure}

In Fig.~\ref{fig:AFM} the GIXRF-reconstructed shape of the sample as well as the AFM profile is overlayed onto the SEM cross section. Therefore the GIXRF line profile and the AFM were scaled to the SEM image in order to match the line pitch while keeping the aspect ratios constant. Also here, the agreement with respect to line height, sidewall angle, line width and even the oxide layer thicknesses (assuming the bright areas in the SEM to be the oxide) is very good. The reconstructed thicker groove oxide can also be seen in the SEM picture.

\section{Conclusion}
Here, we have demonstrated how the GIXRF based methodology for a dimensional and compositional characterization of regular nanostructures can be enhanced with respect to the achieveable sensitivities by incorporating fluorescence signals of different elements from within the nanostructure. In addition, the incorporation of supporting experiments as e.g. XRR for optical constant verification and machine learning techniques such as Bayesian optimization decrease the necessary computational effort of the FEM based reconstruction. The BO allows for an intelligent and fast scanning of the parameter space as compared to other optimizer approaches.

Initial steps towards determining a reliable uncertainty budget for the reconstructed parameter set were taken by deriving confidence intervals for the parameters from the Gaussian process model. We have shown, how the incorporation of the oxygen signal shifts the achievable sensitivities well into the sub-$\si{\nm}$ regime. The obtained GIXRF reconstruction results agree very well with results from SEM and AFM indicating the validity of the methodology.

In addition, we have shown that the methodology is also somewhat sensitive towards unexpected effects on the nanostructure using the example of the carbon contamination. Especially the element sensitivity of X-ray fluorescence but also the behavior of the reconstruction results indicate if unexpected effects are present on the nanostructure. 

By developing more sophisticated techniques to quantify the corresponding model error influences to the final parameter uncertainties this can be a promising technique for nanostructure characterization. In fact, by combining it with techniques such as soft X-ray GISAXS one may even enhance the obtainable sensitivities and even learn about important quality parameters e.g. line roughnesses~\cite{FernandezHerrero2020}.



\section{Acknowledgements}
This project has received funding from the Electronic Component Systems for European Leadership Joint Undertaking under grant agreement No 826589 — MADEin4. This Joint Undertaking receives support from the European Union's Horizon 2020 research and innovation programme and Netherlands, France, Belgium, Germany, Czech Republic, Austria, Hungary, Israel.

JCMwave acknowledges support by the Central Innovation Programm of the Federal Ministry for Economic Affairs and Energy on a basis of a decision by the German Bundestag (project QUPUS, grant no. ZF4450901RR7).

The authors would like to thank Jürgen Probst for producing the samples.





\bibliographystyle{model1-num-names}
\bibliography{sample}







\end{document}